\documentstyle[preprint,floats,prd,aps,eqsecnum,epsf,12pt]{revtex}

\begin{document}
\preprint{\vbox{\hbox{November 1997} \hbox{IFP-748-UNC}
\hbox{Revised May 1998}  }}
\title{\bf Constraints from Precision Electroweak Data on 
Leptoquarks and Bileptons}
\author{\bf Paul H. Frampton and Masayasu Harada}
\address{Department of Physics and Astronomy,
University of North Carolina,\\
Chapel Hill, NC 27599-3255.}
\maketitle

\begin{abstract}
Explicit expressions are derived for 
the oblique parameters $S$ and $T$ 
in certain extensions of the standard model.
In particular, we consider leptoquarks and bileptons, and
find phenomenological constraints on their
allowed masses. 
Leptoquarks suggested by the neutral and charged 
current anomalies at HERA can give improved agreement
with both $S$ and $T$. 
If bileptons are the only new states,
the singly-charged one must be heavier than
the directly-established lower limit.
Finally, we study $SU(15)$ grand unification and show that there are
regions of parameter space where the theory is compatible with 
experimental data.
\end{abstract}

\newpage

\section{Introduction}

Model building for particle theory at and below the TeV
energy scale is not made redundant by the standard
model but is certainly very constrained by it. There
can be little that we do not already know much below the W mass.
In the mass region up to the TeV regime
one expects the Higgs boson and perhaps 
supersymmetric partners. But otherwise one 
must tread softly to avoid upsetting the delicate agreement
between the standard model and precision experimental data.

In the present paper we take a fresh look at
how certain additional bosons can co-exist
with one another and with the successful
standard model. Of special interest are particles
which can contribute negatively to the very useful
parameters $S$ and $T$ which measure compatibility
with precision electroweak data. This is because
these parameters are generally positive
especially for the charged fermionic
particles which one might add.

We shall focus on scalar leptoquarks 
and bileptonic gauge bosons.
Scalar leptoquarks are of special interest becsause of the
experimental anomalies see at HERA (see Section II below).  Bileptons
occur in theoretical frameworks which extend the Standard 
Model (see Section III
below).

The outline of the paper is as follows:
In Section II we treat scalar leptoquarks;
in Section III we analyse bileptons.
In Section IV, we examine whether SU(15) grand
unification can survive its additional mirror fermions.
Finally, in Section V there are our conclusions.

\newpage

\section{Scalar Leptoquarks}
\label{sec:leptoquarks}

The H1~\cite{h1} and ZEUS~\cite{zeus} collaborations reported a
possible excess of neutral current (NC) events in $e^+p$ collisions.
This excess can be explained by the existence of the leptoquark of
mass around $200$\,GeV\cite{lq}.
In addition to the NC channel,
an excess in the charged current (CC) was also
reported\cite{Krakauer}.

Recently, D0~\cite{d0} and CDF~\cite{cdf} gave the 95\% C.L. lower
limit on the masses of 225\,GeV and 213\,GeV with assuming unit
branching fraction into a first-generation charged lepton plus jet.
For compatibility with the HERA anomaly, therefore, the scalar leptoquark
must have a significant branching ratio to other decay channels.

Concerning the contribution from leptoquarks to the weak charge of a nucleus, 
measured by atomic parity violation experiments:
as shown in Ref.~\cite{BKM}, there is a relatively large
contribution coming from tree level diagrams of
$\Delta Q_W^{\rm LQ}$ between $-0.1$ and $-0.8$ which increases the
discrepancy of measurements from the Standard Model
prediction; however, there is non-negligible contribution coming
from the oblique corrections, especially $S$:
$\Delta Q_W^{\rm oblique} \simeq -0.79 S - 0.01 T$~\cite{st}
which could improve agreement for $S$ sufficiently negative.

In Ref.~\cite{BKM} a mixing between two scalar leptoquark multiplets
carrying different weak hypercharges was introduced to simultaneously
explain the NC and CC anomaly at HERA.
Since the lightest leptoquark couples to both $e+j$ and $\nu+j$, the
CDF/D0 limits are weakened.
They also studied the contributions to $\rho$ parameter~\cite{veltman},
and showed that $\Delta\rho$ from leptoquark could be negative.
Since a relatively large mixing is needed, another electroweak
precision parameter $S$~\cite{st} should be also studied.  [The
parameter $T$ is equivalent to $\Delta\rho$.]
The contributions from single SU(2)$_{\rm L}$ doublet of leptoquarks
to $S$ and $T$ are studied in Ref.~\cite{Keith-Ma}.  It is shown that
the contribution to $S$ can be negative while that to $T$ is positive
semidefinite.  However, in the present case, the situation is
different due to the existence of relatively large mixing between two
doublets.

In this section we study the contributions to the oblique parameters
$S$ and $T$ from two doublets of leptoquarks.
We show that both the contributions to $S$ and $T$ can be negative.

The NC and CC anomaly at HERA can be simultaneously explained by
introducing two SU(2)$_{\rm L}$ doublets of leptoquarks with
weak hypercharges $7/6$ and $1/6$~\cite{BKM}.
Let us write these doublets as
$\Phi_{7/6}$ ($Y=7/6$) and $\Phi_{1/6}$ ($Y=1/6$), both of which
belong to a {\bf $3$} representation of SU(3)$_{\rm C}$.
The electric charges are 
$\Phi_{7/6}(Q = 5/3, 2/3)$ and 
$\Phi_{1/6}(Q = 2/3, - 1/3)$. 
The SU(3)$_{\rm C}\times$SU(2)$_{\rm L}\times$U(1)$_{\rm Y}$ invariant
mass terms are given by
\begin{equation}
{\cal L}_M = - M^2 \Phi^{\dag}_{7/6} \Phi_{7/6}
- {M'}^2 \Phi^{\dag}_{1/6} \Phi_{1/6} \ .
\label{inv mass}
\end{equation}
The interactions to the standard Higgs field $H$ are given by
\begin{eqnarray}
{\cal L}_H &=&
- \lambda_1 \left\vert H^{\dag} \Phi_{7/6} \right\vert^2
- \lambda_2 \left\vert H^{\dag} \Phi_{1/6} \right\vert^2
\nonumber\\
&&
- \tilde{\lambda}_1 
  \left\vert \tilde{H}^{\dag} \Phi_{7/6} \right\vert^2
- \tilde{\lambda}_2 
  \left\vert \tilde{H}^{\dag} \Phi_{1/6} \right\vert^2
\nonumber\\
&&
- \lambda_3 \left[
  (\Phi_{7/6})^{\dag} H \tilde{H}^{\dag} \Phi_{1/6}
  + {\rm h.c.}
\right]
\ , 
\label{Higgs}
\end{eqnarray}
where $\tilde{H}=(i\tau_2 H)^T$.\footnote{
There are other terms like 
$\vert H^{\dag} H\vert \Phi_{7/6}^{\dag} \Phi_{7/6}$.  But their
contributions are absorbed into the redefinitions of $M$ and $M'$ in
Eq.~(\ref{inv mass}).
}
After electroweak symmetry breaking by
the vacuum expectation value of $H$,
the $\lambda_1$ and $\tilde{\lambda}_1$ terms give mass splittings
between $Q=5/3$ and $Q=2/3$ components
$(\Phi_{7/6}^{5/3},\Phi_{7/6}^{2/3})$ of $\Phi_{7/6}$, and $\lambda_2$
and $\tilde{\lambda}_2$ terms make mass difference between
$\Phi_{1/6}^{2/3}$ and $\Phi_{1/6}^{-1/3}$.
On the other hand, the $\lambda_3$ term gives mixing between two
$Q=2/3$ leptoquarks of $\Phi_{7/6}$ and $\Phi_{1/6}$.
Let $\varphi$ denote this mixing angle:
\begin{equation}
\left(\begin{array}{c}
  \Phi^{2/3}_l \\ \Phi^{2/3}_h
\end{array}\right)
=
\left(\begin{array}{cc}
  \cos\varphi & -\sin\varphi \\
  \sin\varphi & \cos\varphi  
\end{array}\right)
\left(\begin{array}{c}
  \Phi^{2/3}_{7/6} \\ \Phi^{2/3}_{1/6}
\end{array}\right)
\ ,
\end{equation}
where $\Phi^{2/3}_l$ and $\Phi^{2/3}_h$ denote mass eigenstates.
[We use the convention where the mass of $\Phi^{2/3}_l$ which we
identify with the putative HERA state is lighter
than that of $\Phi^{2/3}_h$.]
We do not write explicit form of mases
$(m_{5/3},m_{-1/3},m_{2/3h},m_{2/3l})$ and the mixing angle $\varphi$
in terms of original parameters in Eqs.~(\ref{inv mass}) and
(\ref{Higgs}), since all of them are independent parameters.  
The D0~\cite{d0} and CDF~\cite{cdf} data imply that the three heavier
leptoquark states should be above 225 GeV if the BR to $e^+q$ is one;
with a smaller BR they can be lighter.
However, it should be
noticed that for the consistency of the model any differences among
all the masses are less than a few 100\,GeV~\cite{BKM}.
Note also that the leptoquark $\Phi_h^{2/3}$ couples to $e^+q$
proportional to $\sin\varphi$ and that as $\varphi$ is increased the
branching ratio $\Phi_l^{2/3}\rightarrow e^+q$ falls below one.

The contributions to $S$ and $T$ from these doublets are given
by~\cite{Lavoura-Li}
\begin{eqnarray}
T &=& \frac{3\sqrt{2}G_F}{16\pi^2 \alpha}
\Biggl[
  \cos^2\varphi 
  \left\{ G_1(m_{5/3},m_{2/3l}) + G_2(m_{2/3h},m_{-1/3}) \right\}
\nonumber\\
&& \qquad
  + \sin^2\varphi
  \left\{ G_1(m_{5/3},m_{2/3h}) + G_2(m_{2/3l},m_{-1/3}) \right\}
\nonumber\\
&& \qquad
  - 4 \sin^2\varphi \cos^2\varphi G_1(m_{2/3l},m_{2/3h})
\Biggr]
\ ,
\nonumber\\
S &=& \frac{3}{\pi}
\Biggl[
  \frac{7}{36} 
  \left\{ 
    \ln \frac{m_{-1/3}^2}{m_{5/3}^2}
    - \cos2\varphi \ln \frac{m_{2/3h}^2}{m_{2/3l}^2}
  \right\}
  - \frac{1}{6} 
  \left\{ 
    \sin^2\varphi \ln \frac{m_{-1/3}^2}{m_{2/3l}^2}
    + \cos^2\varphi \ln \frac{m_{-1/3}^2}{m_{2/3h}^2}
  \right\}
\nonumber\\
&& \qquad
  + \frac{7}{6} 
  \left\{
    G_2(m_Z^2,m_{-1/3},m_{-1/3}) - G_2(m_Z^2,m_{5/3},m_{5/3})
  \right\}
\nonumber\\
&& \qquad
  - \cos2\varphi
  \left\{
    G_2(m_Z^2,m_{2/3h},m_{2/3h}) - G_2(m_Z^2,m_{2/3l},m_{2/3l})
  \right\}
\nonumber\\
&& \qquad
  - \sin^2\varphi
  \left\{
    G_2(m_Z^2,m_{-1/3},m_{-1/3}) - G_2(m_Z^2,m_{2/3l},m_{2/3l})
  \right\}
\nonumber\\
&& \qquad
  - \cos^2\varphi
  \left\{
    G_2(m_Z^2,m_{-1/3},m_{-1/3}) - G_2(m_Z^2,m_{2/3h},m_{2/3h})
  \right\}
\nonumber\\
&& \qquad
  + 2 \sin^2\varphi \cos^2\varphi
  \Bigl\{
    2 G_2(m_Z^2,m_{2/3l},m_{2/3h}) - G_2(m_Z^2,m_{2/3h},m_{2/3h})
\nonumber\\
&& \qquad
    - G_2(m_Z^2,m_{2/3l},m_{2/3l})
  \Bigr\}
\Biggr]
\ ,
\label{eq:ST}
\end{eqnarray}
where $\theta_W$ is the weak mixing angle and
\begin{eqnarray}
G_1(m_1,m_2) &=&
m_1^2 + m_2^2 - \frac{2m_1^2m_2^2}{m_1^2-m_2^2}
\ln \frac{m_1^2}{m_2^2} \ ,
\nonumber\\
G_2(s,M,m) &=&
\bar{F}_3(s,M,m) - \frac{\bar{F}_4(s,M,m)-\bar{F}_4(0,M,m)}{s}
\end{eqnarray}
with $\bar{F}_3$ and $\bar{F}_4$ given in Appendix~\ref{app:A}.

In Figs.~\ref{fig:1}--\ref{fig:3} we show several examples of 
values of $S$ and $T$ with $m_{2/3l}=200$\,GeV and $\varphi=\pi/6$,
$\pi/4$ and $\pi/3$.  For reference, we also show the experimental
fit~\cite{Hagiwara} after subtracting the Standard Model
contribution with Higgs mass of 300\,GeV.
The value of $T$ is more sensitive to the parameter choice than $S$.
Both $S$ and $T$ can be negative for a wide range of parameters.
To compare with experimental data we may need to include vertex
corrections.  However, the Yukawa coupling of leptoquarks to fermions
is small~\cite{lq}, so we expect that vertex and box corrections are
small compared with the contributions through $S$ and $T$.
In such a case leptoquarks can improve the agreement with the data.

\begin{figure}[htbp]
\begin{center}
\ \epsfbox{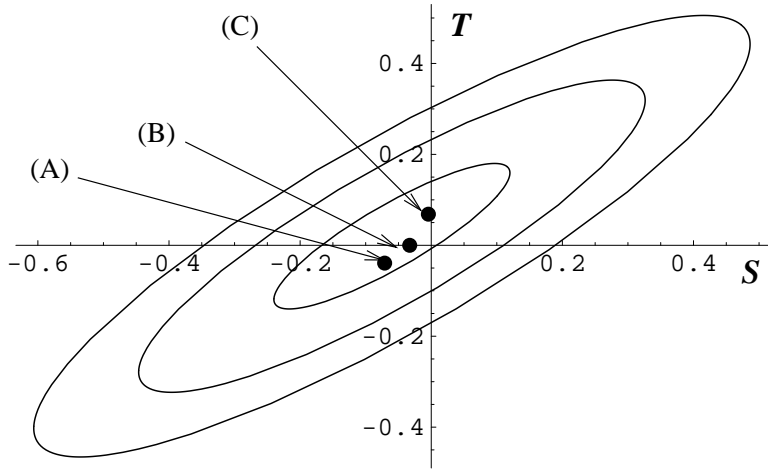}
\end{center}
\caption[]{
Three points indicated by (A), (B) and (C) correspond to the parameter
choices $(m_{2/3h},m_{-1/3},m_{5/3})=(350,345,350)$, $(250,250,250)$
and $(275,300,250)$, respectively.
($m_{2/3l}=200$\,GeV and $\varphi=\pi/4$)
The contours here (and in Figs. 2 and 3) are for 39\%, 90\%,
99\% confidence levels. 
}\label{fig:1}
\end{figure}
\begin{figure}[htbp]
\begin{center}
\ \epsfbox{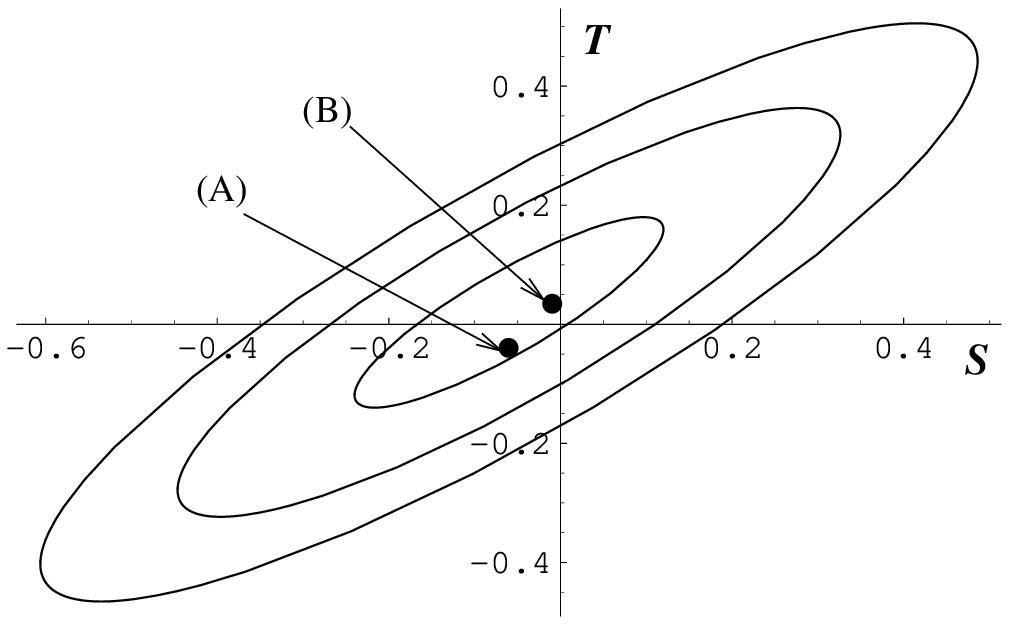}
\end{center}
\caption[]{
The points indicated by (A) and (B) correspond to the parameter
choices $(m_{2/3h},m_{-1/3},m_{5/3})=(310,250,340)$ and
$(250,250,250)$, respectively.
($m_{2/3l}=200$\,GeV and $\varphi=\pi/3$)
}\label{fig:2}
\end{figure}
\begin{figure}[htbp]
\begin{center}
\ \epsfbox{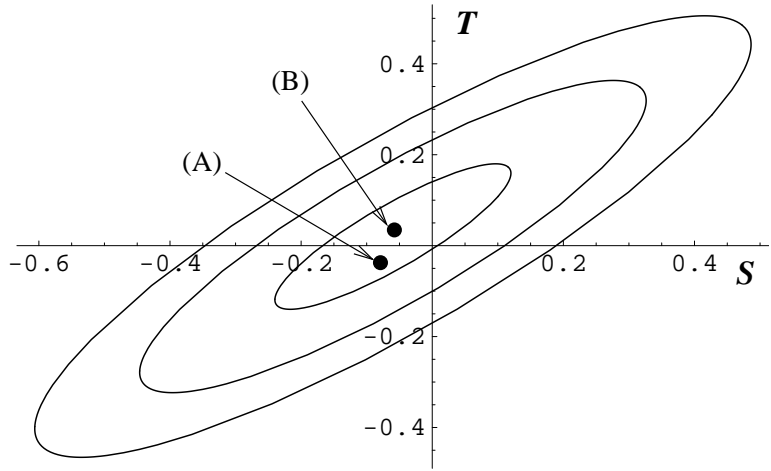}
\end{center}
\caption[]{
The points indicated by (A) and (B) correspond to the parameter
choices $(m_{2/3h},m_{-1/3},m_{5/3})=(300,280,280)$ and
$(250,250,250)$, respectively.
($m_{2/3l}=200$\,GeV and $\varphi=\pi/6$)
}\label{fig:3}
\end{figure}

\newpage

\section{Bileptons}
\label{sec:Bileptons}

In certain extensions of the standard model, there occur
bileptonic gauge 
bosons~\cite{Frampton-Lee,Frampton-Kephart,331} 
which typically occur in
$SU(2)$ doublets $(Y^{--}, Y^-)$ and their conjugates
$(Y^{++}, Y^+)$.
The experimental data currently constrain the masses
of bileptons differently for the singly-charged and doubly-charged
varieties. 
(For a recent review, see, e.g., Ref.~\cite{Cuypers}.)

The best lower limit on the singly-charged
bilepton is presently given by the precision data on
the decay of polarized muons. If the normal decay
$\mu^- \rightarrow e^- + \bar{\nu}_e + \nu_{\mu}$
is contaminated by the bilepton-mediated
$\mu^- \rightarrow e^- + \nu_e + \bar{\nu}_{\mu}$
Fierz rearrangement means that the latter contributes
proportionally to $V + A$ rather than $V - A$.
The limit on the Michel parameter $\xi$ in the 
coupling $V - \xi A$ of $\xi > 0.997$ found
at the PSI experiment\cite{PSI} gives the limit
$M(Y^-) > 230$GeV.

For the doubly-charged bilepton $Y^{--}$ a tighter
lower bound has been found recently from
muonium - antimuonium conversion limits, also at PSI~\cite{muonium}.
The data require that $M(Y^{--}) > 360$GeV.

In the models which predict bileptonic gauge bosons
SU(2)$_{\rm L}\times$U(1)$_{\rm Y}$ is part of a gauged 
SU(3)$_l$, and the bileptonic gauge bosons become massive
when the SU(3)$_l$ is broken.
There are generally two types of models for this kind of bileptons:
(1) bilepton gauge bosons couple to only leptons 
as in SU(15)\cite{Frampton-Lee,Frampton-Kephart};
(2) bilepton gauge bosons couple to quarks as well as leptons 
as in 3-3-1 model\cite{331}.

In the case (1) the usual SU(2)$_{\rm L}$ gauge bosons are certain
linear combinations of gauge bosons of the unbroken SU(2) subgroup of
SU(3)$_l$ and other gauge bosons coupling to quarks.

The generators of SU(2)$_{\rm L}$ and U(1)$_{\rm Y}$,
$I^a$ and $Y$, are embedded as
\begin{equation}
I^a = T_l^a +\cdots \ , \qquad (a=1,2,3) \ , \qquad
Y = -\sqrt{3} T_l^8 + \cdots \ ,
\label{generator relation}
\end{equation}
where $T_l^a$ denote the generators of SU(3)$_l$,
and dots stand for other contributions.
The relations of the gauge bosons $W_\mu^a$ and $B_\mu$ of the standard
SU(2)$_{\rm L}\times$U(1)$_{\rm Y}$ to the $A^a_l$ of SU(3)$_l$
are given by
\begin{eqnarray}
g_l A_\mu^a &=& g W_\mu^a + \cdots \ , \qquad (a=1,2,3) \nonumber\\
g_l A_\mu^8 &=& - \sqrt{3} g' B_\mu + \cdots \ ,
\label{gauge field relation}
\end{eqnarray}
where $g$, $g'$ and $g_l$ denote the corresponding gauge coupling
constants.
The bileptonic gauge bosons are expressed as
\begin{equation}
Y^{\pm\pm}_\mu = \frac{1}{\sqrt{2}}
\left( A_\mu^4 \mp A_\mu^5 \right) \ , \qquad
Y^{\pm}_\mu = \frac{1}{\sqrt{2}}
\left( A_\mu^6 \mp A_\mu^7 \right) \ .
\label{def:bilepton}
\end{equation}

In the models of type (2)
the unbroken SU(2) subgroup of 
SU(3)$_l$ is nothing but the electroweak SU(2)$_{\rm L}$.
Then the first relations of Eqs.~(\ref{generator relation})
and (\ref{gauge field relation}) become
\begin{equation}
I^a = T_l^a \ , \qquad 
g_l A_\mu^a = g W_\mu^a \ , \qquad (a=1,2,3) \ ,
\end{equation}
with $g_l=g$.
The dots parts in the second relations of 
Eqs.~(\ref{generator relation})
and (\ref{gauge field relation}) are modified.
The definitions of bileptonic gauge fields in Eq.~(\ref{def:bilepton})
remain intact.

In both types of models the bileptonic gauge bosons of SU(3)$_l$ makes
SU(2)$_{\rm L}$ doublet with hypercharge 3/2 and its conjugate.
It is convenient to use SU(2)$_{\rm L}$ doublet notation 
$Y_\mu\equiv(Y^{++}_\mu,Y^{+}_\mu)$.
The effective Lagrangian for bileptonic gauge bosons at the scale
below SU(3)$_l$ breaking scale can be written as
\begin{eqnarray}
{\cal L}_{0} &=& - \frac{1}{2} (Y_{\mu\nu})^{\dag} Y^{\mu\nu}
+ (D_\mu \Phi - i M Y_\mu)^{\dag} (D^\mu \Phi - i M Y^{\mu})
\nonumber\\
&& 
- i  g Y_\mu^{\dag} F^{\mu\nu}(W) Y^\mu
+ i \frac{3}{2}  g' Y_\mu^{\dag} F^{\mu\nu}(B) Y^\mu
\ ,
\label{LY0}
\end{eqnarray}
where $\Phi$ are the would-be Nambu-Goldstone bosons eaten by bileptonic
gauge bosons: $\Phi = (\Phi_{++},\Phi_{+})$.
The $Y_{\mu\nu}$ and $D_\mu\Phi$ are given by
\begin{eqnarray}
Y_{\mu\nu} &=&
D_\mu Y_\nu - D_\nu Y_\mu \ ,
\qquad
D_\mu Y_\nu = 
\left[
  \partial_\mu - i g W_\mu + i \frac{3}{2} g' B_\mu
\right]
Y_\nu 
\ ,
\nonumber\\
D_\mu \Phi &=&
\left[
  \partial_\mu - i g W_\mu + i \frac{3}{2} g' B_\mu
\right]
\Phi
\ .
\end{eqnarray}

In the simplest case the SU(2)$_{\rm L}$ doublet Higgs field
is introduced as a part of SU(3)$_l$ triplet (or anti-triplet).
The other component field generally carries lepton number two,
and SU(2)$_{\rm L}$ singlet with hypercharge one or two:
\begin{equation}
\phi = \left(
\begin{array}{c}
H_1 \\ \phi_-
\end{array}
\right)
\ , \qquad
\phi' = \left(
\begin{array}{c}
H_2 \\ \phi_{--}
\end{array}
\right)
\ ,
\end{equation}
where $H_1$ ($H_2$) is a SU(2)$_{\rm L}$ doublet field with 
hypercharge $1/2$ ($-1/2$), and $\phi_-$ ($\phi_{--}$) is a 
SU(2)$_{\rm L}$ singlet field with hypercharge $-1$ ($-2$).

Both the VEVs of these Higgs fields $H_1$ and $H_2$
give masses to $W$ and $Z$ bosons, and the standard electroweak
SU(2)$_{\rm L}\times$U(1)$_{\rm Y}$ is broken.
The VEV of $H_1$ gives a mass correction to $Y^-$, while
the VEV of $H_2$ gives a mass correction to $Y^{--}$.
If only one Higgs doublet had VEV, the mass
difference of bileptons would be related to the mass of $W$ boson.
But in realistic models several Higgs fileds are needed to have VEVs.
In such a case both the masses of bileptons and $W$ boson are
independent with each other.  
In the following we regard the bilepton masses as independent
quantities.
Moreover, the actual would-be Nambu-Goldstone bosons eaten by
bileptonic gauge bosons are certain linear combinations of $\Phi$ in
Eq.~(\ref{LY0}) with $\phi_-$ or $\phi_{--}$.
We assume that the contributions to $S$ and $T$ due to these mixings
are small compared with the bilepton contributions.
Thus we use the following effective Lagrangian for the
kinetic term of would-be Nambu-Goldstone bosons and bilepton masses
after SU(2)$_{\rm L}\times$U(1)$_{\rm Y}$ is broken:
\begin{equation}
{\cal L}_{\rm NG} =
(D_\mu \Phi - i \hat{M} Y_\mu)^{\dag} (D^\mu \Phi - i \hat{M} Y^{\mu})
\ ,
\end{equation}
where $\hat{M}$ is $2\times2$ matrix given by
\begin{equation}
\hat{M} = \left(
\begin{array}{cc}
M_{++} & 0 \\ 0 & M_{+}
\end{array} \right)
\ .
\end{equation}
The calculations below were done in 't Hooft-Feynman gauge.

The contributions to $S$, $T$ and $U$ from bileptonic gauge bosons
coming through the conventional transverse self-energies are given 
by\cite{sasaki}
\begin{eqnarray}
S &=& - 16 \pi {\rm Re} \frac{ \Pi^{3Y}(m_Z^2) - \Pi^{3Y}(0) }{m_Z^2}
\nonumber\\
&=& \frac{9}{4\pi} 
\Biggl[
  \ln \frac{M_{++}^2}{M_+^2}
  + \frac{2}{m_Z^2} 
  \left(
    M_{++}^2 \bar{F}_0(m_Z^2,M_{++},M_{++})
    - M_{+}^2 \bar{F}_0(m_Z^2,M_{+},M_{+})
  \right)
\nonumber\\
&& \qquad
  + \frac{4}{3}
  \left(
    \bar{F}_0(m_Z^2,M_{++},M_{++})
    - \bar{F}_0(m_Z^2,M_{+},M_{+})
  \right)
\nonumber\\
&& \qquad
  - 2
  \left(
    \bar{F}_3(m_Z^2,M_{++},M_{++})
    - \bar{F}_3(m_Z^2,M_{+},M_{+})
  \right)
\Biggr]
\ ,
\nonumber\\
T &=& \frac{4\sqrt{2}G_F}{\alpha}
\left( \Pi^{11}(0) - \Pi^{33}(0) \right)
\nonumber\\
&=&
\frac{3\sqrt{2}G_F}{16\pi^2\alpha}
\left[ 
  M_{++}^2 + M_{+}^2 - \frac{2M_{++}^2M_{+}^2}{M_{++}^2-M_{+}^2}
  \ln \frac{M_{++}^2}{M_{+}^2}
\right]
\ ,
\nonumber\\
U &=& 16\pi \left[
  \frac{\Pi^{11}(m_W^2) - \Pi^{11}(0)}{m_W^2}
  - \frac{\Pi^{33}(m_Z^2) - \Pi^{33}(0)}{m_Z^2}
\right]
\nonumber\\
&=& \frac{1}{2\pi}
\Biggl[
  2 \left(
    \bar{F}_0(m_Z^2,M_{++},M_{++}) + \bar{F}_0(m_Z^2,M_{+},M_{+})
    - 2 \bar{F}_0(m_W^2,M_{++},M_{+})
  \right)
\nonumber\\
&& \qquad
  - 3 \Biggl(
    \frac{M_{++}^2}{m_Z^2} \bar{F}_0(m_Z^2,M_{++},M_{++}) 
    + \frac{M_{+}^2}{m_Z^2} \bar{F}_0(m_Z^2,M_{+},M_{+})
\nonumber\\
&& \qquad\qquad
    - \frac{M_{++}^2+M_+^2}{m_W^2} \bar{F}_0(m_W^2,M_{++},M_{+})
  \Biggr)
\nonumber\\
&& \qquad
  - \frac{3}{m_W^2} \Biggl(
    ( M_{++}^2 - M_+^2 ) 
    \left( 
      \bar{F}_A(m_W^2,M_{++},M_{+}) - \bar{F}_A(0,M_{++},M_{+})
    \right)
\nonumber\\
&& \qquad\qquad
    - ( M_{++}^2+M_+^2 ) \bar{F}_0(0,M_{++},M_{+})
  \Biggr)
\nonumber\\
&& \qquad
  - 3 \left(
    \bar{F}_3(m_Z^2,M_{++},M_{++}) + \bar{F}_3(m_Z^2,M_{+},M_{+})
    - 2 \bar{F}_3(m_W^2,M_{++},M_{+})
  \right)
\Biggr]
\label{STbl:sf}
\end{eqnarray}
where functions $\bar{F}_0$, $\bar{F}_A$ and $\bar{F}_3$ are given in
Appendix~\ref{app:A}.

In the standard model the gauge boson ($W$, $Z$ and photon)
contributions to the $S$, $T$ and $U$ parameters, which are defined in
terms of conventional self-energies, are gauge dependent.
To make these parameters gauge invariant we need to add pinch parts
arising from vertex and box diagrams\cite{PT}.
In the present case the contributions of bilepton gauge bosons to
conventional self-energies are gauge dependent, so that we need to
add pinch parts.

\begin{figure}[htbp]
\begin{center}
\ \epsfbox{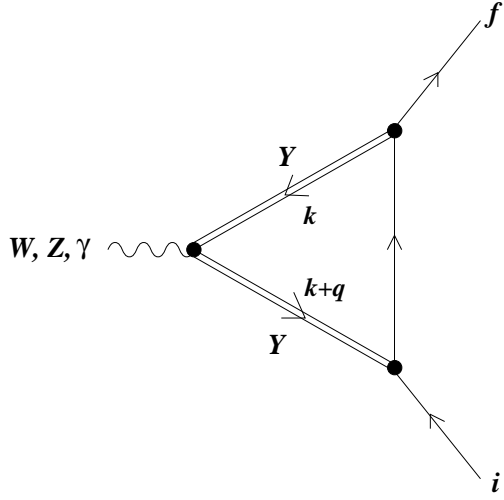}
\end{center}
\caption[]{
Vertex correction which gives pinch contribution.
}\label{fig:pinch}
\end{figure}

In `t Hooft-Feynman gauge this pinch term arises from a vertex
correction which includes a triple vertex of gauge bosons as shown in
Fig.~\ref{fig:pinch}.
We specify here these pinch parts by using current correlation
functions.
Generally the interactions between fermionic currents and 
bileptonic gauge bosons can be expressed as
\begin{equation}
{\cal L} = g_l 
\left[ Y^{++}_\mu J^\mu_{++} + Y^{+}_\mu J^\mu_{+} \right]
+ \mbox{h.c.} \ .
\end{equation}
The $W$-fermion-antifermion vertex correction of Fig.~\ref{fig:pinch}
can be written as
\begin{eqnarray}
\Gamma_{W}^\mu(q) &=&
\frac{g g_l^2}{\sqrt{2}} \int \frac{d^nk}{(2\pi)^2}
\left[
  (-2k-q)_\mu g_{\alpha\beta} + (k+2q)_\alpha g_{\beta\mu}
  + (k-q)_\beta g_{\mu\alpha}
\right]
\nonumber\\
&&
D_{+}(k) D_{++}(k+q) 
\int d^nx e^{ikx}
\langle f \vert T J_{-}^\alpha(x) J_{++}^\beta(0) \vert i \rangle
\ ,
\label{W:vertex}
\end{eqnarray}
where $D_i(k)$ denotes a bilepton propagator denominator:
\begin{equation}
D_i(k) = \frac{1}{k^2-M_i^2} \ .
\end{equation}
The pinch parts are triggered by contractions with the four-momentum
present in the $WYY$ vertex. 
By using the equal time commutator
\begin{equation}
\delta(x_0-y_0) \left[\, J_{++}^0(x) \,,\,J_{-}^\mu(y)\,\right]
= \frac{1}{2} \left( J_{l1}^\mu(x) + i J_{l2}^\mu(x) \right)
\delta^n(x-y) \ ,
\end{equation}
the pinch part of 
Eq.~(\ref{W:vertex}) is expressable as
\begin{equation}
\left. \Gamma_{W}^\mu(q) \right\vert_P = 
g g_l^2 B_0(q^2,M_+,M_{++}) \frac{1}{\sqrt{2}}
\langle f \vert 
\left( J_{l1}^\mu(0) + i J_{l2}^\mu(0) \right)
\vert i \rangle
\ .
\label{W pinch}
\end{equation}
where $B_0$ is defined by
\begin{equation}
B_0(q^2,M_1,M_2) = \int \frac{d^nk}{i(2\pi)^n}
\frac{1}{[M_1^2-k^2][M_2^2-(k+q)^2]} \ .
\end{equation}
By using the relation (\ref{gauge field relation}),
the current associated with the standard SU(2)$_{\rm L}$ gauge boson
$W$ is related to the above $J_{l1}+i J_{l2}$ as
\begin{equation}
\frac{1}{\sqrt{2}}
\left( J_{l1}^\mu + i J_{l2}^\mu \right)
=
\frac{g^2}{g_l^2} J_{W^+}^\mu + \cdots \ .
\label{W current}
\end{equation}
Similarly the pinch part of a $Z$ vertex is given by
\begin{eqnarray}
\left. \Gamma_{Z}^\mu(q) \right\vert_P 
&=&
g^3 
\left( \frac{1-4\sin^2\theta_W}{\cos\theta_W} \right)
B_0(q^2,M_{++},M_{++})
\nonumber\\
&& \quad \times
\left[
  2\sin^2\theta_W \langle f \vert J_Q^\mu \vert i \rangle
  + \frac{1-4\sin^2\theta_W}{2\cos^2\theta_W} 
  \langle f \vert J_Z^\mu \vert i \rangle
\right]
\nonumber\\
&& + g^3 
\left( - \frac{1+2\sin^2\theta_W}{\cos\theta_W} \right)
B_0(q^2,M_{+},M_{+})
\nonumber\\
&& \quad \times
\left[
  \sin^2\theta_W \langle f \vert J_Q^\mu \vert i \rangle
  - \frac{1+2\sin^2\theta_W}{2\cos^2\theta_W} 
  \langle f \vert J_Z^\mu \vert i \rangle
\right]
+ \cdots \ ,
\label{Z pinch}
\end{eqnarray}
and that of a photon vertex by:
\begin{eqnarray}
\lefteqn{
\left. \Gamma_{Q}^\mu(q) \right\vert_P =
4 e g^2 B_0(q^2,M_{++},M_{++})
\left[
  2\sin^2\theta_W \langle f \vert J_Q^\mu \vert i \rangle
  + \frac{1-4\sin^2\theta_W}{2\cos^2\theta_W} 
  \langle f \vert J_Z^\mu \vert i \rangle
\right]
}
\nonumber\\
&& + 2 e g^2 B_0(q^2,M_{+},M_{+})
\left[
  \sin^2\theta_W \langle f \vert J_Q^\mu \vert i \rangle
  - \frac{1+2\sin^2\theta_W}{2\cos^2\theta_W} 
  \langle f \vert J_Z^\mu \vert i \rangle
\right]
+ \cdots \ .
\label{photon pinch}
\end{eqnarray}
The self-energies of electroweak gauge bosons are modified by
pinch parts which can be expressed as
\begin{eqnarray}
\left. \Pi_{ZZ}(q^2) \right\vert_P &=&
(q^2-m_Z^2) 
\Bigl[
  (1-4\sin^2\theta_W)^2 B_0(q^2,M_{++},M_{++})
\nonumber\\
&&\qquad
  + (1+2\sin^2\theta_W)^2 B_0(q^2,M_{+},M_{+})
\Bigr]
\ ,
\nonumber\\
\left. \Pi_{ZQ}(q^2) \right\vert_P &=&
2(2q^2-m_Z^2) (1-4\sin^2\theta_W) B_0(q^2,M_{++},M_{++})
\nonumber\\
&&
-(2q^2-m_Z^2) (1+2\sin^2\theta_W) B_0(q^2,M_{+},M_{+})
\ ,
\nonumber\\
\left. \Pi_{QQ}(q^2) \right\vert_P &=&
4q^2 \left[
  4 B_0(q^2,M_{++},M_{++}) + B_0(q^2,M_{+},M_{+})
\right]
\ ,
\nonumber\\
\left. \Pi_{WW}(q^2) \right\vert_P &=&
2(q^2-m_W^2) B_0(q^2,M_{+},M_{++})
\ .
\label{pinch self energies}
\end{eqnarray}
The corrections to $S$ and $T$ parameters arising from the above pinch
parts are therefore given by
\begin{eqnarray}
\left. S \right\vert_P &=& \frac{1}{\pi}
\Biggl[ 
  3 \ln \frac{M_{++}^2}{M_{+}^2}  
  + 2 ( 1 + 2 \sin^2\theta_W ) \bar{F}_0(m_Z^2,M_{++},M_{++})
\nonumber\\
&& \qquad
  - ( 1 - 4 \sin^2\theta_W ) \bar{F}_0(m_Z^2,M_{+},M_{+})
\Biggr]
\ , 
\nonumber\\
\left. T \right\vert_P &=&
\frac{1}{4\pi\sin^2\theta_W}
\left[
  \frac{M_{++}^2+M_{+}^2}{M_{++}^2-M_{+}^2} 
  \ln\frac{M_{++}^2}{M_+^2}
  - 2 + 3 \tan^2\theta_W \ln\frac{M_{++}^2}{M_+^2}
\right]
\ ,
\nonumber\\
\left. U \right\vert_P &=& \frac{1}{\pi}
\Biggl[ 
  4 \sin^2\theta_W \bar{F}_0(m_Z^2,M_{++},M_{++})
  - 2 \sin^2\theta_W \bar{F}_0(m_Z^2,M_{+},M_{+})
\Biggr]
\ .
\label{SPTP}
\end{eqnarray}
The expressions $S|_P$, $T|_P$ and $U|_P$ of
Eq.(\ref{SPTP}) must be added to $S$, $T$ and $U$ respectively 
of Eq.(\ref{STbl:sf}).

In the models of type (2) such as the 3-3-1 model 
the additional part in Eq.~(\ref{W current}) represented by $\ldots$
does not exist.  So all the pinch part, Eq.~(\ref{W pinch}), is already
included in $\Pi_{WW}(q^2)|_P$ of Eq.~(\ref{pinch self energies}).
The dots parts in Eqs.~(\ref{Z pinch}) and (\ref{photon pinch}) are
proportional to the extra neutral gauge boson ($Z'$) current, so are
related to $Z$--$Z'$ and $\gamma$--$Z'$ mixings.
Similarly in models of type (1) such as SU(15) the $\ldots$ parts in
Eqs.~(\ref{W current})--(\ref{photon pinch}) are related to 
mixings among the electroweak gauge bosons and certain extra gauge
bosons.
These mixings are constrained to be very small.
Then, in the analysis below, instead of introducing new parameters for
these mixings we will neglect the $\ldots$ parts in 
Eqs.~(\ref{W current})--(\ref{photon pinch}).

To compare with the electroweak precision data, we need to include 
the vertex and box corrections as well as
a tree level bilepton contribution to muon decay.
As shown in Ref.~\cite{mudecay,Cuypers}, the singly-charged bilepton
leads to muon decay into $e^-\bar{\nu}_\mu \nu_e$.\footnote{
If bileptons have flavor changing couplings to leptons there are
other decays, but such processes are strongly
constrained~\cite{Cuypers}.
In this paper we assume that bileptons have no flavor-changing
couplings.
}
This decay mode does not interfere with the standard decay (to
$e^-\nu_\mu\bar{\nu}_e$), so the correction to the measured Fermi
constant is given by
\begin{equation}
G_F^{\rm measured} = G_F^{\rm SM} 
\left( 1 + \delta_Y \right)
\label{Fermi constant}
\end{equation}
with
\begin{equation}
\delta_Y = \frac{1}{2} 
\left( \frac{g_l^2}{g^2} \frac{m_W^2}{M_+^2} \right)^2 \ .
\end{equation}
Here, at the tree level of the SM, $G_F^{\rm SM}$ is related to the
mass of $W$ by
\begin{equation}
\left. G_F^{\rm SM}\right\vert_{\rm tree} = 
\frac{g^2}{4\sqrt{2}m_W^2} \ .
\end{equation}
It should be noticed that 
the one-loop correction to the above relation is estimated
as about $0.55\%$ in the SM\cite{Hagiwara}, while
the correction from singly-charged bileptons is
about $\delta_Y\vert\simeq0.005$, $0.0003$ 
for $m(Y+)=250$, $500$\,GeV, respectively.

In models of type (2) bileptons couples to quarks as well as
leptons.  However, such coupling is related to the heavy 
SU(2)$_{\rm L}\times$U(1)$_{\rm Y}$ singlet quarks with lepton number
two.
In the present analysis we consider the case where only the bileptons
are light enough to affect the electroweak precision data.
So we use an effective theory where bileptons couple to only
leptons for both types of models.
The non-pinch part of the $Zl\bar{l}$ vertex correction is then given
by
\begin{eqnarray}
\left. \Gamma_{Zl\bar{l}}(p)^\mu \right\vert_{NP}
&=&
\frac{g}{\cos\theta_W}
\left[
  \Gamma_{Z,L}^l(p^2) \gamma^\mu \frac{1-\gamma_5}{2}
  + \Gamma_{Z,R}^l(p^2) \gamma^\mu \frac{1+\gamma_5}{2}
\right] 
\ ,
\nonumber\\
\Gamma_{Z,L}^l(p^2) 
&=& \frac{g_l^2}{(4\pi)^2}
\left[
  \frac{1-4\sin^2\theta_W}{4} \, \overline{G}_1
    \left(\frac{p^2}{M_{++}}\right)
  + \frac{\sin^2\theta_W}{2} \, \overline{G}_2
    \left(\frac{p^2}{M_{++}}\right)
\right]
\ ,
\nonumber\\
\Gamma_{Z,R}^l(p^2) 
&=& \frac{g_l^2}{(4\pi)^2}
\Biggl[
  \frac{1-4\sin^2\theta_W}{4} \, \overline{G}_1
    \left(\frac{p^2}{M_{++}}\right)
  - \frac{1-2\sin^2\theta_W}{4}\,  \overline{G}_2
    \left(\frac{p^2}{M_{++}}\right)
\nonumber\\
&& \qquad\qquad
  - \frac{1+2\sin^2\theta_W}{4} \, \overline{G}_1
    \left(\frac{p^2}{M_{+}}\right)
  + \frac{1}{4} \, \overline{G}_2
    \left(\frac{p^2}{M_{+}}\right)
\Biggr]
\ ,
\end{eqnarray}
where we have neglected the lepton masses.
The functions $\overline{G}_1$ and $\overline{G}_2$ are given in 
Appendix~\ref{app:A}.
Similarly the non-pinch part of the $Z\nu\bar{\nu}$ vertex correction
is given by
\begin{eqnarray}
\left. \Gamma_{Z\nu\bar{\nu}}(p)^\mu \right\vert_{NP}
&=&
\frac{g}{\cos\theta_W}
\Gamma_{Z,L}^\nu(p^2) \gamma^\mu \frac{1-\gamma_5}{2}
\ ,
\nonumber\\
\Gamma_{Z,L}^\nu(p^2) 
&=& \frac{g_l^2}{(4\pi)^2}
\left[
  - \frac{1+2\sin^2\theta_W}{4} \, \overline{G}_1
    \left(\frac{p^2}{M_{+}}\right)
  + \frac{\sin^2\theta_W}{2} \, \overline{G}_2
    \left(\frac{p^2}{M_{+}}\right)
\right]
\ .
\end{eqnarray}
The non-pinch part of the $W\nu\bar{l}$ vertex correction at
low-energy limit is given by
\begin{eqnarray}
\left. \Gamma_{W\nu\bar{l}}(p)^\mu \right\vert_{NP}
&=&
\frac{g}{\sqrt{2}}
\Gamma_{W,L}(p^2) \gamma^\mu \frac{1-\gamma_5}{2}
\ ,
\nonumber\\
\Gamma_{W,L}(0)
&=&
-\frac{g_l^2}{2(4\pi)^2}
\left[
  \frac{M_{++}^2+M_{+}^2}{M_{++}^2-M_{+}^2}
  \ln \frac{M_{++}}{M_{+}} - 1
\right]
\ .
\end{eqnarray}
Finally the box diagram correction to the above relatin
(\ref{Fermi constant}) is given by
\begin{equation}
\left. \delta_Y \right\vert_{\rm box}
=
\frac{1}{4\sqrt{2}G_F} \frac{g_l^4}{(4\pi)^2}
\left[
  \frac{1}{M_{++}^2-M_{+}^2}
  \ln \frac{M_{++}}{M_{+}}
\right]
\ .
\end{equation}

Combining all the above formulas we have compared the bilepton models
with 15 pieces of data taken from Ref.~\cite{EWdata,Hagiwara} shown
in Table~\ref{tab:EWdata} for convenience; aside from the $W$ mass,
the 14 data are all $Z$-pole quantities.
This implies that the box diagram corrections are needed only for hte
charged current processes.
\begin{table}[htbp]
\begin{tabular}{|l|l|l|}
 & data & SM \\
\hline
$m_Z$ [GeV] & 91.1863 $\pm$ 0.0020 & input \\
$\Gamma_Z$ [GeV] & 2.4946 $\pm$ 0.0027 & 2.4937 \\
$\sigma_{\rm h}^0$ [nb] & 41.508 $\pm$ 0.056 & 41.477 \\
$R_\ell$ & 20.778 $\pm$ 0.029 & 20.736 \\
$A_{\rm FB}^{0, \ell}$ & 0.0174 $\pm$ 0.0010 & 0.0157 \\
${\cal A}_\tau$ & 0.1401 $\pm$ 0.0067 & 0.1439 \\
${\cal A}_e$ & 0.1382 $\pm$ 0.0076 & 0.1438 \\
$R_{\rm b}^0$ & 0.2178 $\pm$ 0.0011 & 0.2157 \\
$R_{\rm c}^0$ & 0.1715 $\pm$ 0.0056 & 0.1721 \\
$A_{\rm FB}^{0,{\rm b}}$ & 0.0979 $\pm$ 0.0023 & 0.1007 \\
$A_{\rm FB}^{0,{\rm c}}$ & 0.0735 $\pm$ 0.0048 & 0.0720 \\
${\cal A}_{\rm b}$ & 0.863 $\pm$ 0.049 & 0.936 \\
${\cal A}_{\rm c}$ & 0.625 $\pm$ 0.084 & 0.667 \\
$A_{\rm LR}$ (SLD) & 0.1542 $\pm$ 0.0037 & 0.1438 \\
$\sin^2\theta_{\rm eff}^{\rm lept}$
($\langle Q_{\rm FB} \rangle$) & 0.2320 $\pm$ 0.0010 & 0.2319 \\
$m_W$ [GeV] & 80.356 $\pm$ 0.125 & 80.324 \\
\end{tabular}
\caption[]{
Electroweak precision data used in the present fit.
The column indicated by SM shows the predictions of the standard model
with $m_H=300$\,GeV.}
\label{tab:EWdata}
\end{table}

Two lower bounds on $m(Y^{++})$ exist in the literature: 360 GeV from 
Ref.~\cite{muonium}, and very recently 850 GeV from
Ref.~\cite{muonium2}.  As well as the best values for $M(Y^{+})$ and
$M(Y^{++})$ we therefore quote the bounds on $M(Y^+)$ for each of the
lower limits on $M(Y^{++})$.

\begin{figure}[htbp]
\begin{center}
\ \epsfbox{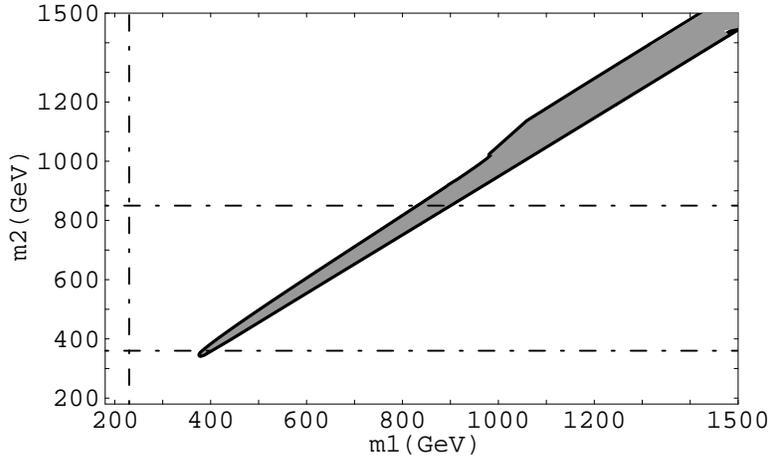}
\end{center}
\caption{
Allowed region (90\%CL) for $m_H = 100\,$GeV.
($m1=m(Y^+)$ and $m2=m(Y^{++})$)
The best overall fit is for masses $m(Y^+) = 566$\,GeV and
$m(Y^{++})=574$\,GeV with $\chi^2/\mbox{d.o.f}=17.6/(15-2)$.
The constraint for $m(Y^+)$ with fixed $m(Y^{++})=360$\,GeV is
$367<m(Y^+)<384$\,(GeV); that with $m(Y^{++})=850$\,GeV is
$790<m(Y^+)<864$\,(GeV).
}
\label{fig:5}
\end{figure}

\begin{figure}[htbp]
\begin{center}
\ \epsfbox{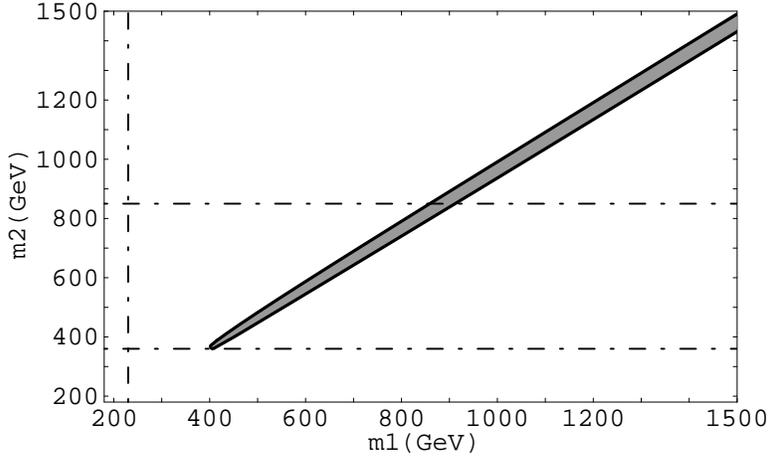}
\end{center}
\caption{
Allowed region (90\%CL) for $m_H = 500\,$GeV.
($m1=m(Y^+)$ and $m2=m(Y^{++})$)
The best overall fit is for masses $m(Y^+) = 1647$\,GeV and
$m(Y^{++})=1602$\,GeV with $\chi^2/\mbox{d.o.f}=17.7/(15-2)$.
$m(Y^{++})=360$\,GeV is excluded.
The constraint for $m(Y^+)$ with fixed $m(Y^{++})=850$\,GeV is
$865<m(Y^+)<905$\,(GeV).
}
\label{fig:7}
\end{figure}

The results for $m_H=100$\,GeV, 300\,GeV and 500\,GeV are respectively
($M(Y^+)$, $M(Y^{++})$)$=$ (566\,GeV, 574\,GeV), (1391\,GeV,
1417\,GeV) and (1647\,GeV, 1602\,GeV).  
(See Figs.~\ref{fig:5}--\ref{fig:7}.)
\begin{figure}[htbp]
\begin{center}
\ \epsfbox{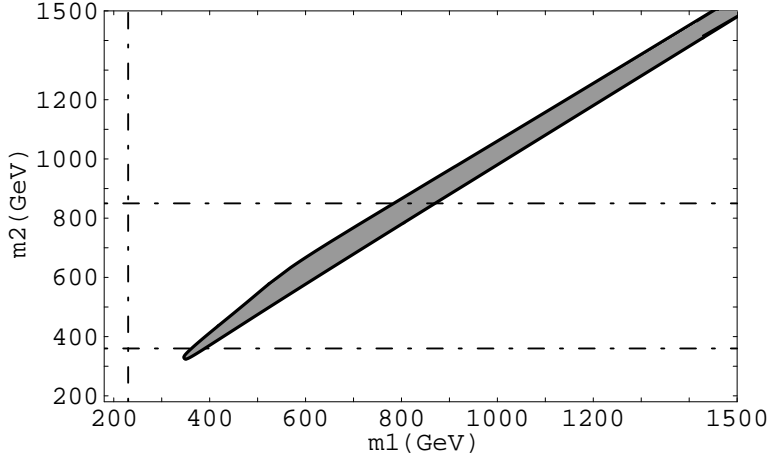}
\end{center}
\caption{
Allowed region (90\%CL) for $m_H = 300\,$GeV.
($m1=m(Y^+)$ and $m2=m(Y^{++})$)
The best overall fit is for masses $m(Y^+) = 1417$\,GeV and
$m(Y^{++})=1391$\,GeV with $\chi^2/\mbox{d.o.f}=17.7/(15-2)$.
The constraint for $m(Y^+)$ with fixed $m(Y^{++})=360$\,GeV is
$387<m(Y^+)<397$\,(GeV);
that with $m(Y^{++})=850$\,GeV is $837<m(Y^+)<894$\,(GeV).
}
\end{figure}
For the fixed values
$M(Y^{++})=$360\,GeV, 850\,GeV the (lower, upper) limits on $M(Y^+)$
are respectively shown in Table~\ref{tab:limit}.
\begin{table}[htbp]
\begin{tabular}{|l|c|c|}
$m_H$ [GeV] & $\left. M(Y^{++})\right\vert_{\rm min}$ [GeV] &
allowed $M(Y^+)$ [GeV] \\
\hline
100 & 360 & $ 367 < M(Y^+) < 384 $ \\
    & 850 & $ 790 < M(Y^+) < 864 $ \\
300 & 360 & $ 387 < M(Y^+) < 397 $ \\
    & 850 & $ 837 < M(Y^+) < 894 $ \\
500 & 360 & excluded \\
    & 850 & $ 865 < M(Y^+) < 905 $ \\
\end{tabular}
\caption[]{
Bounds for $M(Y^+)$ for fixed values of $M(Y^{++})$.}
\label{tab:limit}
\end{table}
In all cases, the lower bound on $M(Y^+)$ is an improvement on the
single-charge bilepton empirical limit (230\,GeV) found in
Ref.~\cite{mudecay}.

\newpage

\section{SU(15) Grand Unification}

In a grand unified model based on SU(15)
~\cite{Frampton-Lee,Frampton-Kephart} each generation of 
quarks and leptons is represented by a fundamental
{\bf 15}.  To cancel anomalies of three generations, three
generations of mirror fermions are needed.  Since these mirror
fermions obtain their masses from the VEV which breaks standard
SU(2)$_{\rm L}\times$U(1)$_{\rm Y}$ symmetry they necessarily
are close to the weak scale in mass and give significant contributions
to $S$ and $T$.
Even if we assume that members of the same SU(2)$_{\rm L}$ doublet have
degenerate masses, and hence the mirror fermions give no contribution
to $T$, they do give a very large contribution
to $S$ parameter; $S_{\rm mirror}=2/\pi$.  Then one might think that
this model is already excluded by the precision electroweak analysis?
However, there are many extra particles including bileptons and
leptoquarks in the model.  These extra particles could give
non-negligible contributions $S$ and $T$ as discussed in the previous
sections.  As is easily read from Eqs.~(\ref{STbl:sf}) and
(\ref{SPTP}), there is a negative contribution to $S$ coming from
bileptons if the singly-charged bilepton ($Y^-$) is heavier than the
doubly-charged one ($Y^{--}$).  This negative contribution can cancel
the large positive contribution coming from mirror fermions.  On the
other hand, such a mass difference of bileptons gives a large
positive contribution to the $T$ parameter.  But this could, in turn,
be canceled by a negative contribution of leptoquarks 
without affecting the $S$ parameter.

The vertex and box corrections are quite negligible in this SU(15)
case where the $Y^{++}$--$Y^{+}$ mass difference is large.  We have
explicitly checked that there exists an extended region of parameter
space where $S$ and $T$ are acceptable: SU(15)
grand unification is not yet excluded by experiment!

\newpage

\section{Conclusions}

The continued robustness of the standard model with respect
to more and more accurate experimental data gives tight
constraints on any attempt at ornamentation of the theory
by additional "light" physics.

The parameters $S$ and $T$ provide a very convenient
measure of compatibility with the precision electroweak data.
Additional states give, in general, postive contributions
to $S$ and $T$ and hence rapidly lead to a possible
inconsistency. It is therefore of special interest to
define what architecture can contribute
negatively to $S$ and $T$.

Here we have discussed two examples: bileptons and 
leptoquarks. For bileptons we have derived a 
lower bound of $367$GeV for the singly-charged
bileptonic gauge boson, assuming that bileptons are the
only states additional to the standard model;
this improves considerably on the mass bound
available from direct measurement.     

If we identify the putative leptoquark at HERA
with a mixture of scalar doublets of SU(2)
having different hypercharge, this can also improve
agreement with data.

Finally we have addressed the exaggerated reports
of the death of SU(15) grand unification due
to the large positive $S$ value from its three 
generations of mirror
fermions. Because of the simultaneous presence of 
both bileptons and 
leptoquarks in SU(15), there is an extended neighborhood
in parameter space where $S$ (and $T$)
can be acceptably small in magnitude.

\section*{Acknowledgement}

This work was supported in part by
the U.S. Department of Energy under
Grant No. DE-FG05-85ER-40219.

\newpage

\appendix

\section{Functions}
\label{app:A}

Here we give several functions used in sections \ref{sec:leptoquarks}
and \ref{sec:Bileptons}.
\begin{eqnarray}
\lefteqn{
  \bar{F}_0(s,M,m) = 
  \int^1_0 dx\, \ln \left( (1-x)M^2 + x m^2 - x(1-x)s \right)
  - \ln M m
}
\nonumber\\
&=&
\left\{
\begin{array}{l}
  -\frac{2}{s} \sqrt{(M+m)^2-s} \sqrt{(M-m)^2-s}
  \ln \frac{ \sqrt{(M+m)^2-s} + \sqrt{(M-m)^2-s} }{ 2 \sqrt{Mm} }
\\
  \qquad
  + \frac{M^2-m^2}{s} \ln \frac{M}{m} - 2 \ ,
  \qquad \mbox{for} \quad s < (M-m)^2 \ ,
\\
  \frac{2}{s} \sqrt{(M+m)^2-s} \sqrt{s-(M-m)^2}
  \tan \sqrt{\frac{s-(M-m)^2}{(M+m)^2-s}}
\\
  \qquad
  + \frac{M^2-m^2}{s} \ln \frac{M}{m} - 2 \ ,
  \qquad \mbox{for} \quad (M-m)^2 < s < (M+m)^2 \ ,
\\
  \frac{2}{s} \sqrt{s-(M+m)^2} \sqrt{s-(M-m)^2}
  \left[
    \ln \frac{ \sqrt{s-(M+m)^2} + \sqrt{s-(M-m)^2} }{ 2 \sqrt{Mm} }
    - i \pi
  \right]
\\
  \qquad
  + \frac{M^2-m^2}{s} \ln \frac{M}{m} - 2 \ ,
  \qquad \mbox{for} \quad (M+m)^2 < s \ ,
\end{array}
\right.
\nonumber\\
\lefteqn{
  \bar{F}_A(s,M,m) =
  \int^1_0 dx\, (1-2x) \ln \left( (1-x)M^2 + x m^2 - x(1-x)s \right)
}
\nonumber\\
&=&
- \frac{M^2-m^2}{s} \left[ \bar{F}_0(s,M,m) - \bar{F}_0(0,M,m) \right]
\ ,
\nonumber\\
\lefteqn{
  \bar{F}_3(s,M,m) = 
  \int^1_0 dx\, x(1-x)\ln \left( (1-x)M^2 + x m^2 - x(1-x)s \right)
  - \frac{1}{6}\ln M m
}
\nonumber\\
&=&
\frac{1}{6} \left[ 
  1 + \frac{M^2+m^2}{s} - \frac{2(M^2-m^2)^2}{s^2}
\right]
\bar{F}_0(s,M,m)
\nonumber\\
&& \quad
-\frac{1}{6} \left( 1 - \frac{2(M^2+m^2)}{s} \right)
\frac{M^2-m^2}{s} \ln \frac{M}{m} + \frac{1}{18}
- \frac{(M^2-m^2)^2}{3s^2} \ ,
\nonumber\\
\lefteqn{
  \bar{F}_4(s,M,m) = 
  \int^1_0 dx\, \left[ (1-x)M^2+xm^2 \right] \,
  \ln \left( (1-x)M^2 + x m^2 - x(1-x)s \right)
}
\nonumber\\
  && \qquad
  - \frac{M^2+m^2}{2}\ln M m
\nonumber\\
&=&
  \frac{M^2+m^2}{2} \bar{F}_0(s,M,m) +
  \frac{M^2-m^2}{2} \bar{F}_A(s,M,m)
\ .
\end{eqnarray}
The functions used in the vertex corrections for bileptons are given 
by
\begin{eqnarray}
\overline{G}_1(a) &=&
\frac{2(1+2a)}{a(4-a)} \left[ \left(J(a)\right)^2 + a J(a) \right]
+ J(a) + \frac{9a}{2(4-a)}
\ , 
\nonumber\\
\overline{G}_2(a) &=&
\frac{7}{2} + \frac{2}{a} - \left( 3 + \frac{2}{a} \right) \ln ( -a )
+ 2 \left( 1 + \frac{1}{a} \right)^2
\left[ \mbox{Sp}(-a) + \ln ( -a ) \ln ( 1 + a ) \right]
\ ,
\end{eqnarray}
where $J(a)=\bar{F}_0(a,1,1)$ and $\mbox{Sp}(x)$ is the Spence's
function defined by
\begin{equation}
\mbox{Sp}(x) = - \int^x_0 \frac{dt}{t}\, \ln ( 1- t )
\ .
\end{equation}

\end{document}